\documentclass[prl,reprint,twocolumn,showpacs,superscriptaddress,floatfix,aps,10pt]{revtex4-2}

\usepackage{amsmath,amsthm,amssymb}
\usepackage{dcolumn,bm,hyperref}
\usepackage{graphicx,subfigure,verbatim}
\usepackage{newtxtext}
\usepackage{xcolor}
\usepackage{soul}
\usepackage{ulem}
\usepackage{graphicx}
\usepackage{braket}

\usepackage{xcolor}

\begin{document}

\title{Quantum Spin Transfer of Spin-Correlated Electron Pairs}
\author{Seongmun Hwang}
\affiliation{Department of Physics, Korea Advanced Institute of Science and Technology (KAIST), Daejeon 34141, Korea}
\author{Jung Hyun Oh}
\affiliation{Department of Physics, Korea Advanced Institute of Science and Technology (KAIST), Daejeon 34141, Korea}
\author{Paul M. Haney}
\affiliation{Physical Measurement Laboratory, National Institute of Standards and Technology, Gaithersburg, Maryland 20899, USA}
\author{Mark D. Stiles}
\affiliation{Physical Measurement Laboratory, National Institute of Standards and Technology, Gaithersburg, Maryland 20899, USA}
\author{Kyung-Jin Lee}
\email{kjlee@kaist.ac.kr}
\affiliation{Department of Physics, Korea Advanced Institute of Science and Technology (KAIST), Daejeon 34141, Korea}

\date{\today}

\begin{abstract}
We theoretically investigate quantum spin transfer from spin-correlated conduction-electron pairs to localized spins in a ferromagnet, given that electrons are correlated intrinsically. We show that even spin-singlet pairs and triplet pairs with $m=0$, both carrying no net spin, can transfer finite angular momentum through the quantum fluctuation term inherent to the $sd$ exchange interaction. The amount of transferred spin differs between the singlet and triplet $m=0$ states due to quantum interference. The difference is such that the independent-electron approximation remains valid for spin transfer when injected spin currents are completely incoherent. However, in partially coherent systems, like superconductor/ferromagnet junctions, coherent spin-singlet currents can directly convert into equal-spin triplet currents in generic ferromagnets, without requiring magnetic inhomogeniety or spin-orbit coupling. 
\end{abstract}

\maketitle

{\it Introduction.}---Spin transfer~\cite{slonczewski1996,berger1996,ralph2008}, the transfer of spin angular momentum from conduction electrons to localized magnetic moments, is a central phenomenon in nonequilibrium spin physics. It also underpins key spintronic functionalities such as current-driven magnetization switching~\cite{myers1999,katine2000,zhang2002,mangin2006}, oscillation~\cite{tsoi1998,kiselev2003,chen2016}, and domain-wall motion~\cite{zhang2004,Yamaguchi2004,Yamanouchi2004,Klaui2005,parkin2008}. 
Conventional spin transfer theory~\cite{slonczewski1996,berger1996} treats the localized spin as a classical vector of fixed magnitude, whose dynamics follow the Landau-Lifshitz-Gilbert equation with a spin-transfer torque \(\boldsymbol{\tau}_{\rm STT}\propto\mathbf{m}\times(\mathbf{m}\times\braket{\hat{\boldsymbol{\sigma}}})\), where \(\mathbf{m}\) is the unit vector along the magnetization, corresponding to the localized spin, and \(\braket{\hat{\boldsymbol{\sigma}}}\) is the expectation value of conduction electron spin. Under this classical assumption for localized spins, only the transverse component of the incoming spin relative to \(\mathbf{m}\) can be absorbed and exert torque on the magnetization~\cite{zhang2002,stiles2002}.

Several prior studies have revealed fundamental limitations of this classical assumption. A cryogenic spin-valve experiment showed that the injection of minority spins enhances quantum fluctuations of the magnetization, resulting in highly nonclassical states~\cite{zholud2017}. This phenomenon, arising from the quantum nature of localized spins, is termed {\it quantum spin transfer}. In contrast to the conventional picture, the quantum treatment of localized spins enables finite spin transfer even from longitudinal spin currents polarized antiparallel to the local spins~\cite{mondal2019,mitrofanov2021,petrovic2021}. This quantum spin transfer effect becomes more pronounced in atomic-scale magnets such as magnetic adatoms~\cite{Loth2010,Alexander2013} and molecular magnets~\cite{Kovarik2024}. Moreover, longitudinal spin pumping—the Onsager reciprocal of quantum spin transfer—was recently observed during the antiferromagnet-to-ferromagnet phase transition of FeRh, where the Rh moments undergo large-amplitude longitudinal dynamics, i.e., a temporal change in the atomic moment's magnitude~\cite{lee2025}. These findings call for a re-examination of the classical treatment of localized spins in conventional theory.

Prior studies of quantum spin transfer treat both the conduction and localized spins quantum mechanically~ \cite{mondal2019,mitrofanov2021,petrovic2021,kim2007,mitrofanov2020,Petrovic2021b,Suresh2024}, but they commonly neglect the intrinsic correlation among conduction electrons. In reality, electrons are always correlated, and any spin current is necessarily carried by such correlated electrons. Understanding how these spin-correlated electrons transfer angular momentum to a quantum magnet is therefore essential, and it also enables an assessment of the commonly used independent-electron approximation, in which each electron is assumed to move independently in an effective potential that incorporates correlation effects. While this approximation is justified for charge transport~\cite{Slater1951}, its validity for spin transport remains unclear.

In this paper, we investigate quantum spin transfer from a spin-correlated electron pair to a ferromagnetic local spin chain. Our main findings are twofold. First, finite spin transfer occurs even when the incoming electron pair carries zero net spin, as in the spin singlet or triplet $m=0$ states, where $m$ is the magnetic quantum number—an effect absent in the conventional theory. Second, due to quantum interference, the spin singlet and triplet $m=0$ states yield different transfer magnitudes despite both having identical (zero) spin expection values. The differences are such that when taken together, they demonstrate that the independent-electron approximation is valid for spin transfer when injected spin currents are completely incoherent. Finally, we discuss how the spin singlet state can directly transition to the equal-spin triplet state (i.e., $m=1$) in generic ferromagnets,
which enables long-range propagation of supercurrents in the ferromagnet.

{\it Model.}---Although electron correlations can involve more than two electrons, we focus on the injection of a conduction-electron pair consisting of two electrons prepared in an entangled state as the simplest case. 
We begin by defining the Hilbert space of the composite system, which consists of the conduction-electron pair and a chain of localized spins:
\begin{equation} \label{eq1}
    \mathcal{H} = \mathcal{H}^{\rm orb}_A\otimes\mathcal{H}^{\rm orb}_B\otimes\mathcal{H}^{\rm spin}_{\rm el}\otimes\mathcal{H}^{\rm spin}_{\rm loc},
\end{equation}
where \(\mathcal{H}^{\rm orb}_{A(B)}\) denotes the orbital subspace of electron $A(B)$ in the pair, \(\mathcal{H}^{\rm spin}_{\rm el}\) is the spin subspace of the pair, and \(\mathcal{H}^{\rm spin}_{\rm loc}\) is the spin subspace of the localized spin chain. We take a single $s$-orbital per site, giving \(\dim{\mathcal{H}^{\rm orb}_{A(B)}}=N\), where \(N\) is the number of atomic sites. All spins are taken as $1/2$, so that \(\dim{\mathcal{H}^{\rm spin}_{\rm el}}=4\) and \(\dim{\mathcal{H}^{\rm spin}_{\rm loc}}=2^L\), with \(L\) the number of localized spins.

We construct the following tight-binding Hamiltonian~\cite{mondal2019,mitrofanov2021,petrovic2021,kim2007,mitrofanov2020}:
\begin{widetext}
\begin{eqnarray} \label{eq2}
{\mathcal H} =&& -t_h \sum^{N-1}_{i=1}\left[\ket{i}\bra{i+1}_A\otimes\mathbb{I}+\mathbb{I}\otimes\ket{i}\bra{i+1}_B+H.c.\right] -J_{\rm ex}\sum^{L-1}_{j=1} \boldsymbol{\hat{S}}^{(j)}\cdot\boldsymbol{\hat{S}}^{(j+1)} \nonumber\\
&&-J_{\rm sd}\sum^{L}_{j=1} \left[\ket{n_j}\bra{n_j}_A\otimes\mathbb{I}\otimes\boldsymbol{\hat{\sigma}}^{(A)}\cdot\boldsymbol{\hat{S}}^{(j)}+\mathbb{I}\otimes\ket{n_j}\bra{n_j}_B\otimes\boldsymbol{\hat{\sigma}}^{(B)}\cdot\boldsymbol{\hat{S}}^{(j)}\right],
\end{eqnarray}
\end{widetext}
where \(\mathbb{I}\) is the identity operator, \(n_j\) denotes the atomic site of the \(j^{th}\) local spin, \(\boldsymbol{\hat{S}}^{(j)}\) is the corresponding local spin operator, and \(\boldsymbol{\hat{\sigma}}^{(A)}\) and \(\boldsymbol{\hat{\sigma}}^{(B)}\) are the spin operators of conduction electrons $A$ and $B$.
The first term represents nearest-neighbor hopping with hopping constant \(t_h\). The second term is the ferromagnetic exchange between localized spins, with \(J_{\rm ex}>0\). The third is the $sd$ exchange between  conduction and localized spins, with \(J_{sd}>0\), which mediates spin transfer. To eliminate boundary effects, we impose periodic boundary conditions. 

\begin{figure}[b]
    \centering
    \includegraphics[width=8.6cm]{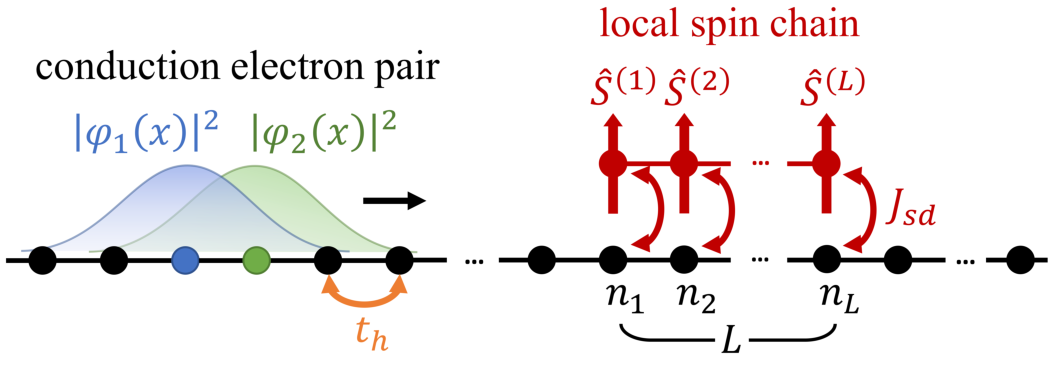}
    \caption{Schematic illustration of the model. A conduction-electron pair is initially placed on the left side of the spin chain. Each electron is described by a Gaussian wave packet, \(\varphi_1(x)\) and \(\varphi_2(x)\), sharing the same wavevector \(k_e\) and width \(\sigma_d\). Their centers are separated by a distance \(d\). As time evolves, the pair traverses the spin chain. The hopping constant is \(t_h\). Localized spins occupy sites \(n_1\) to \(n_L\) and interact with the conduction spins through the $sd$ exchange interaction with coupling strength \(J_{sd}\). This model generalizes the single-electron injection considered in Ref.~[\onlinecite{mitrofanov2021}] to the injection of a spin-correlated electron pair.}
    \label{fig:1} 
\end{figure}

The initial state is a direct product of the conduction electron state and the local spin chain state. The conduction electron state, consisting of orbital and spin sectors, satisfies antisymmetry under particle exchange as required by the Pauli exclusion principle. For the spin sector, we consider either the spin singlet state or one of the spin triplet states. The orbital sector is a symmetric or antisymmetric superposition of Gaussian wave packets, 
\begin{equation}
\ket{\varphi_1}_A\ket{\varphi_2}_B\pm\ket{\varphi_2}_A\ket{\varphi_1}_B,
\end{equation}
where 
\begin{equation}
\braket{x|\varphi_j}=\frac{1}{\sqrt{\sigma_d\sqrt{\pi}}}\exp\left(ik_e(x-x_j)-\frac{(x-x_j)^2}{2\sigma_d^2}\right)
\end{equation}
for \(j=1,2\), where \(k_e,\sigma_d\) and \(x_j\) denote the electron wave vector, packet width, and packet center, respectively. We set \(x_2>x_1\) so that \(\ket{\varphi_2}\) enters the spin chain before \(\ket{\varphi_1}\). A schematic illustration of the model is shown in Fig.~\ref{fig:1}.

The local spin chain is initialized in the ferromagnetic ground state with all spins \(\uparrow\) along the $z$-axis, and the conduction electron spins are aligned collinearly. This configuration restricts dynamics to the spin-$z$ component, while the transverse components remain strictly zero. The time evolution of the total state \(\ket{\Psi (t)}\) is obtained by numerically solving the Schr\"odinger equation. Spin expectation values are evaluated from the density matrix, \(\rho(t)=\ket{\Psi(t)}\bra{\Psi(t)}\). The net spin-$z$ components of the conduction electron pair and the local spin chain are respectively given by
\begin{eqnarray}
\langle\hat{\sigma}_z(t)\rangle&=&\mathrm{Tr}\left[\rho^{\rm spin}_{\rm el}(t)\left(\hat{\sigma}^{(A)}_z+\hat{\sigma}^{(B)}_z\right)\right], \\
\langle\hat{S}_z(t)\rangle&=&\sum^L_{i=1}\mathrm{Tr}\left[\rho^{\rm spin}_{\rm loc}(t)\hat{S}_z^{(i)}\right],
\end{eqnarray}
where \(\rho^{\rm spin}_{\rm el}(t)\) and \(\rho^{\rm spin}_{\rm loc}(t)\) are the reduced density matrices for the spaces \(\mathcal{H}^{\rm spin}_{\rm el}\) and \(\mathcal{H}^{\rm spin}_{\rm loc}\), respectively. The spin transfer is quantified by the change in the conduction electron spin-$z$ component from its initial value, \(\Delta\sigma_z=|\langle\hat{\sigma}_z(t)\rangle - \langle\hat{\sigma}_z(0)\rangle|\). 

\begin{figure}[b]
    \centering
    \includegraphics[width=8.6cm]{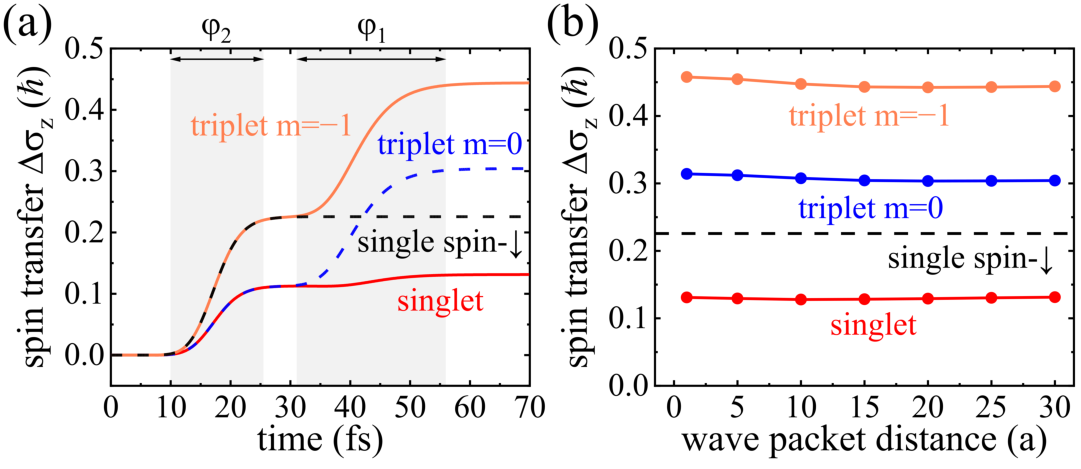}
    \caption{(a) Time evolution of the spin transfer. The red solid line corresponds to the singlet state, blue dotted line to the triplet $m=0$ state, orange solid line to the triplet $m=-1$ state, and black dotted line to a single spin-\(\downarrow\) state. The wave-packet distance is set to 30\(a\), where \(a\) denotes the atomic spacing. The shaded regions represent the time intervals during which $\varphi_1$ and $\varphi_2$ stay within the spin chain, respectively. (b) Spin transfer as a function of the wave-packet distance for the singlet, triplet $m=0$, and triplet $m=-1$ states. Parameters: \(t_h=0.5, J_{ex}=0.2, J_{sd}=0.3\) (in units of eV), \(N=200, L=4, k_e=1/a\) and \(\sigma_d=5a\).}
    \label{fig:2} 
\end{figure}

{\it Results.}---Figure~\ref{fig:2}(a) shows the time evolution of the spin transfer \(\Delta\sigma_z\) for a single spin-\(\downarrow\) state and for three types of electron pairs: the singlet, triplet $m=0$, and triplet $m=-1$ states. \(\Delta\sigma_z\) becomes nonzero at around 10 fs, when  one of the wave packets enters the ferromagnetic chain. The triplet $m=1$ state is not shown because it is an eigenstate of the ferromagnet and therefore exhibits no spin transfer. For all three electron-pair states, \(\Delta\sigma_z\) displays a characteristic two-step structure in time, arising from the sequential entry of the two spatially separated wave packets, while the single spin-\(\downarrow\) state exhibits only one step. The net spin transfer for each state can be estimated from the second plateau, corresponding to the time at which both packets have passed through the local spin chain. Throughout the entire evolution, $\langle\hat{\sigma}_z(t)\rangle+\langle\hat{S}_z(t)\rangle$ remains constant (not shown), confirming conservation of the total spin-$z$ component due to the spin-$z$ rotational symmetry, \([\hat{\sigma}_z+\hat{S}_z,{\mathcal H}]=0\). 

As expected, the triplet $m=-1$ state yields a spin transfer roughly twice that of a single spin-\(\downarrow\) state, reflecting the fact that it carries two \(\downarrow\)-spins. Figure~\ref{fig:2}(a), however, reveals two nontrivial observations: (I) finite spin transfer occurs even for electron pairs carrying zero net spin, such as the singlet and triplet $m=0$ states, and (II) these two states exhibit different spin-transfer magnitudes despite having identical (zero) spin expection values.

We first discuss observation (I). The finite spin transfer from the singlet and triplet $m=0$ states stands in sharp contrast to the conventional theory, which predicts no transfer in the absence of spin polarization in the injected electrons, thereby highlighting its intrinsically quantum origin. Additional support for its quantum nature comes from the nonlocal character of these correlated states. Because the singlet and triplet $m=0$ states are spatially nonlocal two-electron states, we examine how their nonlocality influences spin transfer by varying the spatial separation $d$ between the two wave packets [Fig.~\ref{fig:2}(b)]. We find that both states show nearly constant spin-transfer magnitudes over a wide range of $d$. This insensitivity to the wave-packet separation demonstrates that the spin transfer itself is nonlocal—a hallmark of its quantum mechanical origin.

The origin of observation (I) can be understood by rewriting the $sd$ exchange interaction as,
\begin{equation} \label{eq7}
\hat{H}_{sd} = -J_{sd} \left[\hat{\sigma}_z^{(\alpha)}\hat{S}_z^{(j)}+\frac{1}{2}\left(\hat{\sigma}_{+}^{(\alpha)}\hat{S}_-^{(j)}+\hat{\sigma}_-^{(\alpha)}\hat{S}_+^{(j)}\right)\right],
\end{equation}
where \(\hat{\sigma}_\pm^{(\alpha)}\) and \(\hat{S}_\pm^{(j)}\) denote the spin ladder operators for the conduction electron spin with \(\alpha=A,B\), and the \(j^{th}\) localized spin, respectively. For the collinear spin configurations considered here, the second term of Eq.~(\ref{eq7}), $(\hat{\sigma}_{+}\hat{S}_-+\hat{\sigma}_-\hat{S}_+)$, generates longitudinal quantum fluctuations and enables spin transfer between the conduction and localized spins~\cite{mitrofanov2021,petrovic2021,lee2025}, while the first term, $\hat{\sigma}_z\hat{S}_z$, cannot transfer spin. 

\begin{figure}
    \centering
    \includegraphics[width=8.6cm]{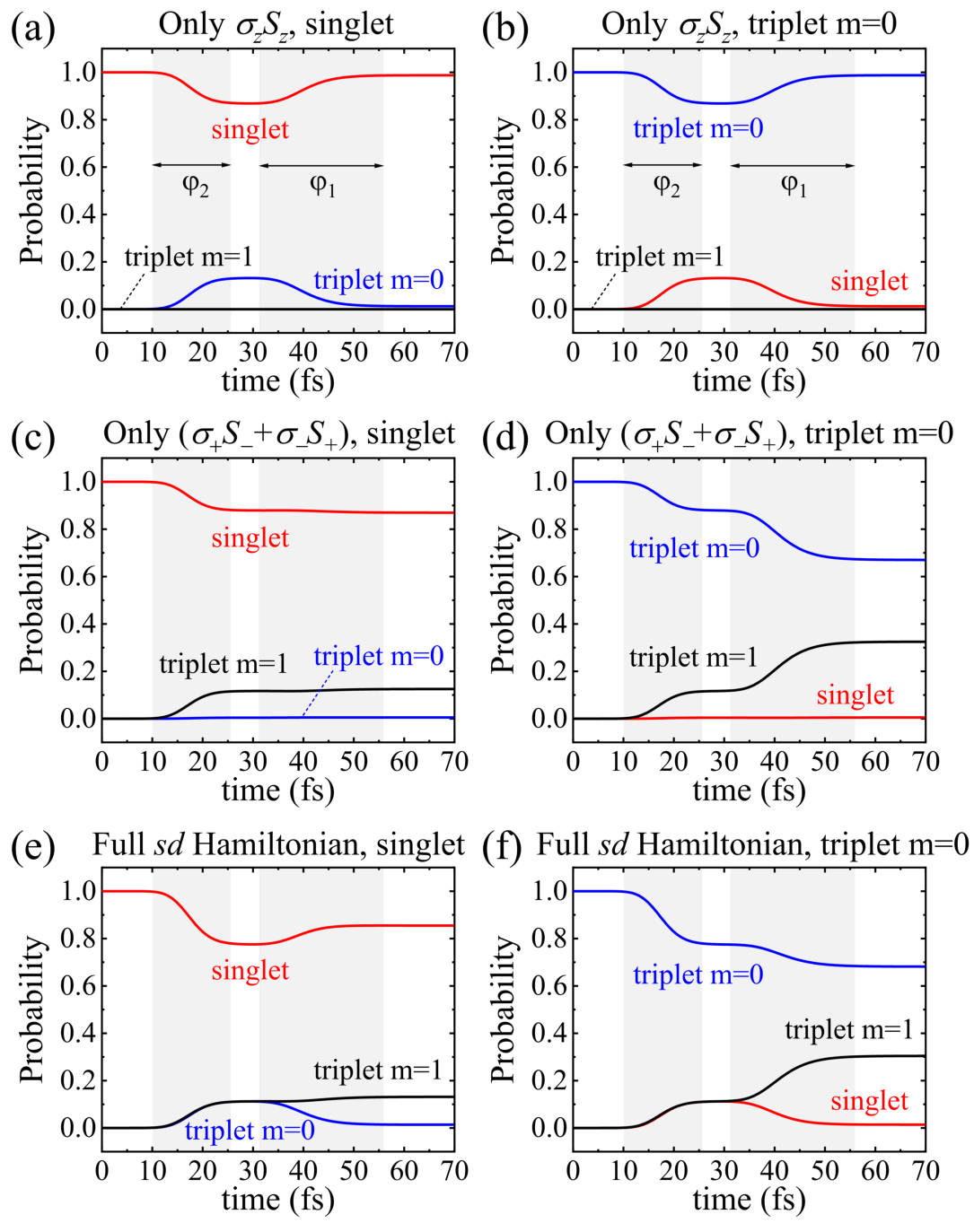}
    \caption{Time evolution of the spin-state probabilities as the conduction electron pair traverses the spin chain, with the initial spin state chosen as the singlet state (left column) or the triplet m=0 state (right column). The shaded regions represent the time intervals during which $\varphi_1$ and $\varphi_2$ stay within the spin chain, respectively.  Panels (a)-(b) include only the \(\hat{\sigma}_z\hat{S}_z\) term, panels (c)-(d) include only the $(\hat{\sigma}_{+}\hat{S}_-+\hat{\sigma}_-\hat{S}_+)$ term, and panels (e)-(f) show the results for the full Hamiltonian in Eq.(\ref{eq7}).}
    \label{fig:3} 
\end{figure}

To verify this, we calculate the time evolution of the transition probabilities from the singlet and triplet $m=0$ states to other states while retaining either the first or the second term of the $sd$ exchange interaction in Eq.~(\ref{eq7}) [Fig.~\ref{fig:3}]. When only the first term is retained, the injected singlet  partially transitions to the triplet $m=0$ state [Fig.~\ref{fig:3}(a)] and the injected triplet $m=0$ state partially transitions to the singlet [Fig.~\ref{fig:3}(b)]. Since both states carry zero net spin, the first $\hat{\sigma}_z\hat{S}_z$ term produces no spin transfer to the localized spins, but merely mixes the two states. This behavior reflects the fact that the spin-$z$ operator does not flip spins but only induces spin-dependent phase shifts. 
We note that the singlet-to-triplet $m=0$ and triplet $m=0$-to-singlet transition probabilities vanish once the two wave packets leave the ferromagnetic region [i.e., after 50 fs in Fig.~\ref{fig:3}(a) and (b)]. It is because the mixing occurs only in the ferromagnet without causing any spin change and thus ceases outside the interaction region. 

In contrast, when only the second term is retained, both the singlet and triplet $m=0$ states develop finite transition probabilities to the triplet $m=1$ state.
This arises because the second term with spin ladder operators mediates simultaneous spin flips of the conduction and localized spins. The singlet (triplet $m=0$) state is an antisymmetric (symmetric) coherent superposition of the \(\ket{\uparrow\downarrow}_{\rm el}\) and \(\ket{\downarrow\uparrow}_{\rm el}\). When either electron occupies a local-spin site, the ladder operators flip the corresponding local spin and drive the transition of conduction-electron state to \(\ket{\uparrow\uparrow}_{\rm el}\); the transition to \(\ket{\downarrow\downarrow}_{\rm el}\) is forbidden by conservation of the spin-$z$ component. Because the triplet $m=1$ state carries finite spin, these transitions directly contribute to spin transfer, confirming that the second $(\hat{\sigma}_{+}\hat{S}_-+\hat{\sigma}_-\hat{S}_+)$ term is responsible for the effect. Importantly, the triplet $m=1$ component, once created, remains finite even after the wave packets have passed the spin chain, since these transitions correspond to real spin-flip processes that transfer angular momentum to the localized spins. As a side remark, a weak mixing between the singlet and triplet $m=0$ states may occur through higher-order processes involving the \(\ket{\uparrow\uparrow}_{\rm el}\) state, but this contribution is small in this example. Indeed, the singlet-to-triplet $m=0$ transition probability in Fig.~\ref{fig:3}(c) is barely visible compared with the dominant singlet-to-triplet $m=1$ transition, and the same tendency appears in Fig.~\ref{fig:3}(d). Figures~\ref{fig:3}(e) and (f) show the evolution of the transition probabilities under the full $sd$ exchange Hamiltonian, which is given by the combined contributions of the first and second terms.

We next discuss observation (II): the different spin-transfer magnitudes obtained from the injection of the singlet and triplet $m=0$ states, despite their identical (zero) spin expection values. While both states are coherent superpositions of \(\ket{\uparrow\downarrow}_{\rm el}\) and \(\ket{\downarrow\uparrow}_{\rm el}\), they differ in their relative phases—$\pi$ for the singlet and $0$ for the triplet $m=0$. This relative phase leads to difference in quantum interference. In the transition to \(\ket{\uparrow\uparrow}_{\rm el}\) state, the singlet state undergoes destructive interference, whereas the triplet $m=0$ state experiences constructive interference. Consequently, these interference effects suppress (for the singlet) or enhance (for the triplet $m=0$) the amount of spin transfer relative to that of a single spin-$\downarrow$ electron.

\begin{figure}
    \centering
    \includegraphics[width=8.6cm]{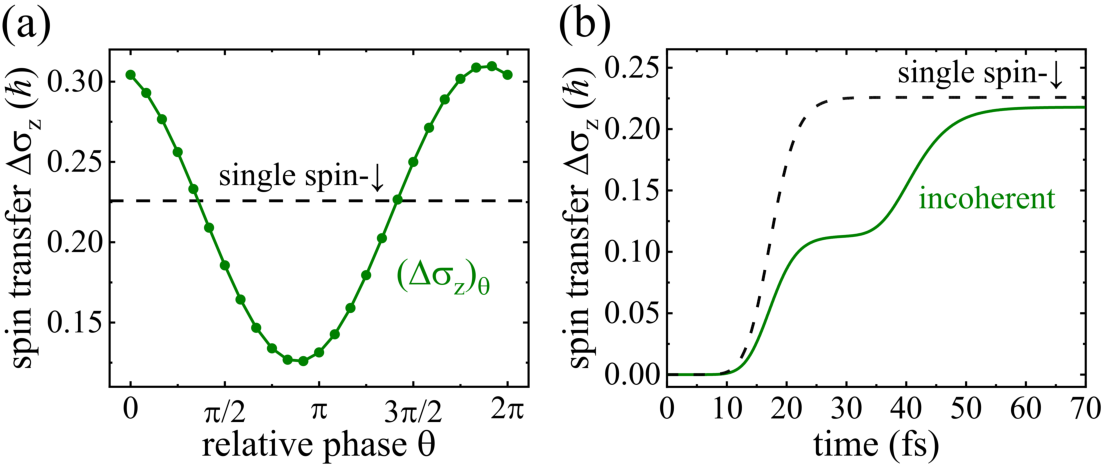}
    \caption{(a) Spin transfer as a function of the relative phase $\theta$. The incoming state is given in Eq.(\ref{eq9}), and $\theta$ is varied from 0 to \(2\pi\). (b) Time evolution of the spin transfer for an incoherent spin state, obtained by averaging \((\Delta\sigma_z)_\theta\) over \(\theta\). The spin transfer of a single spin-\(\downarrow\) is shown as a black dotted line for comparison.}
    \label{fig:4} 
\end{figure}

To support this, we examine how the spin transfer depends on the relative phase of the initial state. A conduction-electron pair in which \(\ket{\uparrow\downarrow}_{\rm el}\) and \(\ket{\downarrow\uparrow}_{\rm el}\) are coherently superposed with a relative phase \(\theta\) can be written as
\begin{eqnarray} \label{eq9}
    \ket{\Psi}_\theta = &\ket{\varphi_1}_A\ket{\varphi_2}_B\otimes\frac{1}{\sqrt{2}}\left(\ket{\uparrow\downarrow}_{\rm el}-e^{i\theta}\ket{\downarrow\uparrow}_{\rm el}\right) \nonumber
    \\-&\ket{\varphi_2}_A\ket{\varphi_1}_B\otimes\frac{1}{\sqrt{2}}\left(\ket{\downarrow\uparrow}_{\rm el}-e^{i\theta}\ket{\uparrow\downarrow}_{\rm el}\right).
\end{eqnarray}
Using \(\ket{\Psi}_\theta\) as the initial state, we compute the spin transfer \((\Delta\sigma_z)_\theta\) as a function of \(\theta\) [Fig.~\ref{fig:4}(a)]. We find that the spin transfer exhibits a clear sinusoidal dependence on the relative phase (see Supplementary Materials~\cite{SM_PRL2025} for a small phase shift). Since the net spin momentum remains zero for all \(\theta\), this result demonstrates that the spin transfer is governed not only by net spin polarization of injected electrons but also by their quantum interference.

The above result suggests that electron correlations play an essential role in determining the amount of spin transfer as long as the spin state preserves phase coherence: in this regime, the electrons cannot be treated as independent. It also implies that when many wave functions with random phases are superposed so that the state becomes incoherent, the resulting spin transfer should approach that of a single spin-\(\downarrow\), since the interference contributions average out to zero. To evaluate the spin transfer of an incoherent state, we average \((\Delta\sigma_z)_\theta\) over the relative phase \(\theta\in [0, 2\pi]\). Figure~\ref{fig:4}(b) shows the corresponding time evolution, demonstrating that the spin transfer indeed converges to that of a single spin-\(\downarrow\) state once all wave packets have passed through the spin chain (see Supplementary Materials~\cite{SM_PRL2025} for a small difference in the spin transfer between the incoherent state and the single spin-\(\downarrow\) state). This confirms that the additional contribution arising from quantum interference vanishes for a completely incoherent spin state, and the spin transfer is then determined solely by the number of spin-\(\downarrow\) electrons. Therefore, electrons can be treated independently when the spin state is fully incoherent (see Supplementary Materials~\cite{SM_PRL2025} for the detailed condition under which the independent-electron approximation holds); otherwise, the independent-electron approximation breaks down for spin transfer.

{\it Discussion.}---We show that quantum spin transfer from intrinsically correlated electron pairs reflects nonlocality and phase coherence. The dependence of spin transfer on the relative phase demonstrates that quantum coherence of the injected electrons plays a central role in determining angular-momentum transfer.

Our findings also clarify when the independent-electron approximation becomes valid for spin transfer. Because interference contributions vanish when the injected spins are completely incoherent, the relevant condition is set by the competition between the phase-coherence time of the longitudinal spin, $\tau_\phi$, and the injection interval, $\tau_{\rm inj}$. When $\tau_{\rm inj}\gg\tau_\phi$, injected spins lose their phase coherence before the next one arrives, resulting in random relative phases; under this condition, the spin transfer reduces to that of a single spin-$\downarrow$ electron, and the independent-electron approximation holds. When $\tau_{\rm inj}\le \tau_\phi$, phase coherence is preserved and electron correlations must be treated explicitly.

In realistic spin-orbit torque experiments on normal metal/ferromagnet bilayers, typical charge current densities of $j_c\approx5\times10^{11}$ A/m$^2$ correspond to an injection interval per atomic site of $\tau_{\rm inj}\approx e/(\theta_{\rm SH} j_c a^2)\approx10$ ps, assuming a spin Hall angle $\theta_{\rm SH}\approx0.3$ and a lattice constant $a\approx0.3$ nm. Although the longitudinal spin phase-coherence time $\tau_\phi$ of correlated electron pairs is unknown, it must be limited by the longitudinal spin relaxation time $\tau_{\rm sf}$, which sets the upper bound of $\tau_\phi$. Given $\tau_{\rm sf}\approx2$ ps for Co~\cite{Stiles2004}, $\tau_{\rm inj}\gg\tau_\phi$ is likely satisfied in typical spin-orbit torque experiments, implying that conduction electron spins are effectively incoherent and justifying the independent-electron approximation. 

In contrast, the condition $\tau_{\rm inj}\gg\tau_\phi$ may be substantially relaxed in atomic-scale magnets~\cite{Loth2010,Alexander2013,Kovarik2024} where only a few localized spins participate and the spin coherence can exceed that of thin-film ferromagnets. A further implication arises for superconducting spintronics~\cite{linder2015}. Unlike normal metal/ferromagnet structures, where singlet and triplet $m=0$ states are injected with same weight (thus, effectively incoherent), superconductor/ferromagnet junctions inject a coherent singlet state into the ferromagnet. In this case, a singlet Cooper pair can directly convert into a triplet $m=1$ state inside the ferromagnet through quantum spin transfer, generating equal-spin triplet currents. Such equal-spin triplet currents enable long-range supercurrent propagation in the ferromagnet~\cite{buzdin2005,eschrig2011,linder2015,cai2023}. Crucially, this singlet-to-triplet $m=1$ conversion occurs without requiring magnetic inhomogeneity or spin–orbit coupling, in contrast to conventional mechanisms for inducing triplet correlations in superconductor/ferromagnet systems~\cite{bergeret2001,khaire2010,robinson2010,anwar2012,bergeret2014,eskilt2019}. This mechanism therefore provides a simple and potentially robust route for generating equal-spin triplet supercurrents in hybrid superconductor/ferromagnet devices. 

In summary, our work shows that quantum coherence and electron correlations are not merely corrections but can fundamentally dictate spin transfer in magnets whenever spin coherence is well preserved. These findings advance the microscopic understanding of spin–angular-momentum flow and offer promising pathways toward coherent spintronics and superconducting triplet technologies.

{\it Acknowledgments.}---We thank Sergei Urazhdin, Satoru Nakatsuji, Matthew W. Daniels, and Dan Gopman for fruitful comments. This work was supported by the National Research Foundation of Korea (NRF) grant funded by the Korea government (RS-2024-00436660, RS-2024-00487645, RS-2024-00410027).

\bibliographystyle{apsrev4-2}
\bibliography{references}


\end{document}